\documentclass[aps,prl,twocolumn,showpacs,superscriptaddress,floatfix]{revtex4}

\usepackage{amsmath}
\usepackage{amssymb}
\usepackage{amsmath}
\usepackage{amsbsy}
\usepackage{graphicx}
\usepackage{amsfonts}%
\usepackage{graphicx}
\usepackage{epsfig}
\usepackage[english]{babel}

\newcommand{\Eq}[1]{Eq.~(\ref{#1})}

\newcommand{\be}{\begin{equation}}
\newcommand{\bea}{\begin{eqnarray}}
\newcommand{\eea}{\end{eqnarray}}
\newcommand{\ee}{\end{equation}}

\newcommand{\qqph}{\qquad \phantom{.}}
\newcommand{\MPS}{{{\rm MPS}}}
\newcommand{\NRG}{{{\rm NRG}}}

\def\one{\ensuremath{\hbox{$\mathrm I$\kern-.6em$\mathrm 1$}}}

\begin{document}


\title{Variational matrix product state approach to quantum impurity models}

\author{A. \surname{Weichselbaum}}
\affiliation{
   Physics Department, Arnold Sommerfeld Center for Theoretical
   Physics, and Center for NanoScience,
   Ludwig-Maximilians-Universit\"at M\"unchen, 80333 M\"unchen,
   Germany
}

\author{F. \surname{Verstraete}}
\affiliation{
   Institut der Theoretischen Physik, Universit\"at Wien,
   Boltzmanngasse 3, A-1090 Wien, Austria
}

\author{U. \surname{Schollw\"ock}}
\affiliation{
   Institut f\"ur Theoretische Physik C, RWTH Aachen University,
   D-52056 Aachen, Germany
}

\author{J. I. \surname{Cirac}}
\affiliation{
   Max-Planck-Institut f\"ur Quantenoptik,
   Hans-Kopfermann-Str. 1, Garching, D-85748, Germany
}

\author{Jan \surname{von Delft}}
\affiliation{
   Physics Department, Arnold Sommerfeld Center for Theoretical
   Physics, and Center for NanoScience,
   Ludwig-Maximilians-Universit\"at M\"unchen, 80333 M\"unchen,
   Germany
}

\date{May 14, 2008}

\begin{abstract}
We present a unified framework for renormalization group methods,
including Wilson's numerical renormalization group (NRG) and White's
density-matrix renormalization group (DMRG), within the language of
matrix product states. This allows improvements over Wilson's NRG for
quantum impurity models, as we illustrate for the one-channel Kondo
model. Moreover, we use a variational method for evaluating Green's
functions. The proposed method is more flexible in its description
of spectral properties at finite frequencies, opening the way to
time-dependent, out-of-equilibrium impurity problems. It also
substantially improves computational efficiency for one-channel
impurity problems, suggesting potentially \emph{linear} scaling of
complexity for $n$-channel problems.
\end{abstract}


\maketitle

Wilson's numerical renormalization group (NRG) is a key method
\cite{Wilson} for solving quantum impurity models such as the Kondo,
Anderson or spin-boson models, in which a local degree of freedom,
the ``impurity'', is coupled to a continuous bath of excitations.
These models are of high relevance in the description of magnetic
impurities, of quantum dots, and problems of decoherence. NRG has
been used with great success to calculate both thermodynamic
\cite{Wilson,Krishna} and dynamical \cite{Costi,Hofstetter,Raas}
properties. It is, however, of limited use in more complex
situations: Computational cost grows exponentially for a coupling to
multiple bands in the bath. In systems out of equilibrium or with
time-dependent external parameters, such as occur in the tuning of
quantum dots, difficulties arise due to NRG's focus on low energy
properties through its logarithmic discretization scheme which looses
accuracy at high spectral frequencies.

In the present Letter, we draw attention to the fact that
states generated by the NRG have the structure of \emph{matrix
product states} (MPS) \cite{AKLT+Fannes,VPC}
on a 1-D geometry. This is a simple observation, which however has
important conceptual and practical implications:

   (i) As White's density matrix renormalization group
(DMRG) \cite{White} for treating quantum chain models is
in its single-site version identical to variational MPS
\cite{VPC}, NRG and DMRG are now seen to have the same
formal basis of matrix product states, resolving a
long-standing question about the connection between both
methods.
   (ii) \emph{All} NRG results can be improved upon
systematically by \emph{variational optimization} in the
space of variational matrix product states (VMPS) of the
same structure as those used by NRG. This does not lead
to major improvements at $\omega=0$ where NRG works very
well, but leads to the inclusion of feedback from low-
to high-energy states, also allowing the relaxation of
the logarithmic bath discretization of NRG: spectra away
from $\omega=0$ can be described more accurately and
with higher resolution.
   (iii) Recent algorithmic advances using VMPS
\cite{VPC}, in particular those treating time-dependent
problems \cite{t-methods,VGC}, can now be exploited to
tackle quantum impurity models involving time-dependence
or nonequilibrium; this includes applications to the
description of driven qubits coupled to decohering
baths, as relevant in the field of quantum computation.
   (iv) The VMPS algorithm allows ground state properties
of quantum impurity models to be treated more
efficiently than NRG: the same accuracy is reached in
much smaller ansatz spaces (roughly, of square-root
size). Moreover, our results suggest that for many (if
not all) $n$-channel impurity problems it should be
feasible to use an \emph{unfolded} geometry, for which the
complexity will only grow linearly with $n$.

The present Letter provides a ``proof of principle'' for the VMPS
approach to quantum impurity models by applying it to the
one-channel Kondo model. We reproduce the NRG flow of the finite
size spectrum \cite{Krishna}, and introduce a VMPS approach for
calculating Green's functions, as we illustrate for the impurity spectral
function \cite{Costi}, which yields a significant improvement over
existing alternative techniques \cite{GreensDMRG1, GreensDMRG2,
GreensDMRG3, othermethods}. Our results illustrate in which sense
the VMPS approach is numerically more efficient than the NRG.

\emph{NRG generates matrix product states:---} To be specific, we
consider Wilson's treatment of the Kondo model, describing a local
spin coupled to a fermionic bath. To achieve a separation of energy
scales, the bath excitations are represented by a set of
logarithmically spaced, discrete energies $\omega_n = \Lambda^{-n}$,
where $\Lambda > 1$ is a ``discretization parameter'' \cite{Wilson}.
By tridiagonalization, the model is then mapped onto the form of a
semi-infinite chain $\mathcal{H} = \lim_{N \to \infty}
\mathcal{H}^N$ where \cite{Wilson}
\begin{equation}
   \mathcal{H}^N = -2J \mathbf{s} \cdot \mathbf{S} +
   \sum_{n=1}^{N-1} \xi_{n}\left( c_{n\mu}^{\dagger} c_{n+1,\mu} +
   c_{n+1,\mu}^{\dagger}c_{n\mu}\right)
\label{eq.Hamilton}
\end{equation}%
with $\mathbf{S} \equiv \textstyle{\frac{1}{2}}
c_0^\dagger \mathbf{\boldsymbol\sigma} c_0$
and creation (annihilation) operators $c^\dagger\,(c)$,
respectively.
$\mathcal{H}^N$ describes an impurity spin $\mathbf{s}$ coupled
to the first site of a chain of length $N$ of Fermions with
spin $\mu$ and exponentially decreasing hopping matrix elements
along the chain ($\xi_n \sim \Lambda^{-n/2}$).
$\mathcal{H}^N$ lives on a Hilbert space spanned by
the set of $d_I d^N$ basis states $\{ |i_0, i_1, i_2, \dots i_N
\rangle\}$, where $i_0$ labels the $d_I$ possible impurity states
and $i_n$ (for $n = 1, \dots, N$) the $d$ possible states of site $n$
(for the Kondo model, $i_0=\left\{ \uparrow, \downarrow \right\}$
and for all other sites $i_n = \left\{ 0, \uparrow, \downarrow,
\uparrow\downarrow \right\}$, i.~e. $d_I=2$ and $d=4$).

To diagonalize the model, NRG starts with a chain of length
$(\bar n-1)$, chosen sufficiently small that $\mathcal{H}^{\bar
n -1}$ can be diagonalized exactly, yielding a set of
eigenstates $|\psi_\alpha^{\bar n-1}\rangle$. One continues
with the subsequent iterative prescription: project
$\mathcal{H}^{\bar n -1}$ onto the subspace spanned by its
lowest $D$ eigenstates, where $D < d_I d^{\bar n -1}$ is a
control parameter (typically between 500 and 2000); add site
$\bar n$ to the chain and diagonalize $\mathcal{H}^{\bar n}$ in
the enlarged $ (Dd)$-dimensional Hilbert space, writing the
eigenstates as
\begin{eqnarray}
   |\psi^{\bar n }_\beta \rangle =
   \sum_{i_{\bar n} = 1}^d \sum_{\alpha= 1}^D |\psi^{\bar
   n-1}_\alpha \rangle |i_{\bar n}\rangle
   \, P^{[i_{\bar n}]}_{\alpha \beta } \; ,
\label{eq:NRGrepresentation}
\end{eqnarray}
where the coefficients have been arranged in a matrix
$P^{[i_{\bar n}]}_{\alpha \beta}$ with matrix indices $\alpha,\beta$,
labelled by the site index $\bar n$ and state index $i_{\bar n}$;
rescale the eigenenergies by a factor $\Lambda^{1/2}$; and repeat,
until the eigenspectrum converges, typically for chain lengths $N$
of order 40 to 60.  At each step of the iteration, the eigenstates
of $\mathcal{H}^N$ can thus be written [by repeated use of
\Eq{eq:NRGrepresentation}] in the form of a so-called \emph{matrix
product state},
\begin{eqnarray}
   |\psi^{N}_{\alpha}\rangle &=&
   P^{[i_{0}]}_{\alpha_0 }
   P^{[i_{1}]}_{\alpha_0\alpha_1}
   P^{[i_{2}]}_{\alpha_1\alpha_2}\ldots
   P^{[i_{N}]}_{\alpha_{N-1}\alpha} |i_0, i_1, \dots, i_N
   \rangle  \qqph \label{MPS1}
\end{eqnarray}
(summation over repeated indices implied). The ground state is
then the lowest eigenstate of the effective Hamiltonian
$\mathcal{H}^N_{\alpha \beta} = \langle \psi^N_\alpha |
\mathcal{H}^N | \psi^N_\beta \rangle$, i.e.\ the projection of
the original $\mathcal{H}$ on the subspace of MPS of the form
(\ref{MPS1}).

\emph{VMPS optimization:---} Let us now be more ambitious, and aim
to find the \emph{best possible} description of the ground state
within the space of all MPSs of the form (\ref{MPS1}), using the
matrix elements of the matrices $\{ P^{[n]} \}$ with $P^{[n]} \equiv
\{ P^{[i_n]} \}$ as \emph{variational parameters} to minimize the
energy.  Using a Lagrange multiplier to ensure normalization, we thus
study the following optimization problem:
\begin{eqnarray}
   \min_{|\psi^N\rangle\in\{ {\rm MPS} \}} \left[ \langle \psi^N |
   \mathcal{H}^N | \psi^N \rangle -\lambda\langle\psi^N|\psi^N
   \rangle \right].
\end{eqnarray}
This cost function is multiquadratic in the $d_I+d(N-1)$ matrices
$\{P^{[n]}\}$ with a multiquadratic constraint. Such problems can be
solved efficiently using an iterative method in which one fixes all
but one (let's say the $\bar n$'th) of the matrices $\{P^{[n]}\}$
at each step; the optimal $P^{[\bar n]}$ minimizing the cost
function given the fixed values of the other matrices can then be
found by solving an eigenvalue problem \cite{VPC}. With $P^{[n]}$
optimized, one proceeds the same way with $P^{[\bar{n}+1]}$ and so
on. When all matrices have been optimized locally, one sweeps back
again, and so forth. By construction, the method is guaranteed to
converge as the energy goes down at every step of the iteration,
having the ground state energy as a global lower bound.  Given the
rather monotonic hopping amplitudes, we did not encounter problems
with local minima.

In contrast, NRG constructs the ground state in a single
\emph{one-way} sweep along the chain: each $P^{[n]}$ is thus
calculated only once, without allowing for possible feedback of
$P$'s calculated later. Yet viewed in the above context, the ground
state energy can be lowered further by MPS optimization sweeps. This
accounts for the \emph{feedback} of information from low to high
energy scales. This feedback may be small in practice, but it is not
strictly zero, and its importance increases as the logarithmic
discretization is refined by taking $\Lambda \to 1$.  Note that the
computational complexity of both VMPS optimization and NRG scales as
$Nd D^3$ \cite{White,VPC}, and symmetries can be exploited (with
similar effort) in both approaches. The inclusion of feedback leads
to a better description of spectral features at high frequencies,
which are of importance in out-of-equilibrium and time-dependent
impurity problems. Moreover, it also allows to relax the logarithmic
discretization scheme, further improving the description of
structures at high frequency as illustrated below.

\emph{Energy level flow:---} The result of a converged set of
optimization sweeps is a VMPS ground state $|\tilde
\psi_0^N\rangle$ of the form (\ref{MPS1}); exploiting a gauge
degree of freedom \cite{VPC}, the $\tilde P$'s occurring therein
can always be chosen such that all vectors $|\tilde
\psi_{\alpha}^n\rangle = \bigl[ \tilde P^{[i_0]} \dots \tilde
P^{[i_n]}\bigr]_\alpha |i_0, \dots, i_n\rangle$ are
orthonormal.  The effective Hamiltonian at chain length $n$,
the central object in NRG, is then
$  \tilde{\mathcal{H}}^n_{\alpha \beta} =  \langle
     \tilde{\psi}^n_\alpha | \Lambda^{n/2}\mathcal{H}^n | \tilde{\psi}^n_\beta
   \rangle
$.
Its eigenspectrum can be monitored as $n$ increases,
resulting in an energy level flow along the chain.

\emph{Green's functions:---} Similar techniques also
allow Green's functions to be calculated variationally
\cite{othermethods}. The typical Green's functions of
interest are of the form
$G_\eta^c(\omega)=\langle\psi_0|c |\chi \rangle$ where
$|\chi\rangle$, commonly called a correction vector
\cite{Soos}, is defined by
\begin{eqnarray}
|\chi\rangle \equiv \frac{1}{\omega-\mathcal{H}+i\eta} c^\dagger
|\psi_0\rangle \; , \label{eq:correctionvector}
\end{eqnarray}
with $|\psi_0\rangle$ the ground state of the system, e.~g.
calculated using the VMPS approach and thus represented as MPS.
The spectral density is then given by ${\cal A} (\omega) =
-\lim_{\eta\rightarrow 0} \frac{1}{\pi}
\Im\mathrm{m}\left(G_\eta^c(\omega)\right)$.
The (unnormalized) state $|\chi \rangle$ may be calculated
\emph{variationally} within the set of MPS by optimizing the
weighted norm
\begin{eqnarray}
\mathcal{N}=\left\Vert\chi\rangle-\frac{1}{\mathcal{H}-\omega-i\eta}
c^\dagger|\psi_0\rangle \right\Vert_{W=\left(\mathcal{H}-
\omega\right)^2+\eta^2} , \label{norm}
\end{eqnarray}
where $\Vert\xi\rangle\Vert_W^2 \equiv \langle\xi|W|\xi\rangle$, and
weight $W>0$ such that it yields a quadratic equation.
Minimizing $\mathcal{N}$ efficiently by optimizing one $P$
at a time leads to two independent optimizations
over $\Re\mathrm{e}|\chi\rangle$ and $\Im\mathrm{m}|\chi\rangle$, respectively.
Both involve only multilinear terms such that each
iteration step requires to solve a sparse linear
set of equations \cite{VGC}.

The minimization of ${\cal N}$ in \Eq{norm}, however, can
become increasingly ill-conditioned as $\eta \rightarrow 0$
\cite{EPAPS}, with conditioning deteriorating quadratically in
$\eta$. If one directly solves $\delta / \delta \langle
P^{[n]} | \ \left[ \langle \chi | (\mathcal{H}-\omega-i\eta)
|\chi\rangle - \langle \chi | c^\dagger |\psi\rangle \right] \equiv 0$
by a non-hermitian equation solver such as the
biconjugate gradient method, conditioning deteriorates only linearly.

\begin{figure}[t]
\includegraphics[width=1\linewidth]{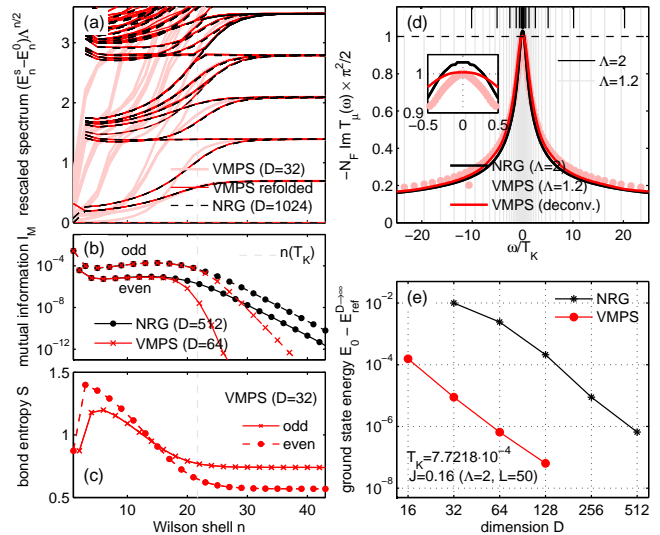}
\caption{
      (Color online) Comparison of VMPS and NRG data for
   logarithmic discretization. (a) Energy level flow of the
   Kondo model as a function of site index $n$ obtained from ${\cal
   H}^{\rm eff}_\mu$ of a variationally optimized MPS with
   $D_\MPS=32$ (light red), of the corresponding recombined spin
   chains (red) \cite{Saberi08}, and from NRG using $D_\NRG = 32^2$
   states (dashed black).
      (b) Correlation along the Wilson chain between
   spin-up and spin-down at site $n$ in terms of mutual information
   $I_M (n) \equiv S({n_\uparrow}) + S({n_\downarrow}) -
   S(n_\uparrow,n_\downarrow)$. Here $S$ is the entropy of the
   reduced density matrix of the groundstate with respect to the
   indicated subspace \cite{Saberi08} (solid for even, dashed for
   odd sites $n$). The Wilson shell corresponding to $T_K$ is
   indicated by the vertical dashed line.
      (c) Bond entropy $S$ along the unfolded Wilson chain
   where $S$ is the usual von Neumann entropy of the VMPS
   reduced density matrix when going from site $n$ to $n+1$.
      (d) Comparison of $T$-matrix ($\Im\mathrm{m}\,T_\mu$, see
   also Fig.~\ref{Fig_Spectral}) for $B=0$ between VMPS and NRG,
   including deconvoluted VMPS data \cite{EPAPS}.  Inset shows zoom
   into peak at $\omega=0$. The significantly smaller $\Lambda=1.2$
   applicable for VMPS (discretization intervalls are indicated by
   vertical lines) shows clearly improved agreement with the Friedel
   sum rule $T\left(0\right)\,\pi^2/2=1$.
      (e) Comparison of ground state energy of the Kondo Hamiltonian
   (\ref{eq.Hamilton}) for fixed chain length relative to the
   extrapolated energy for $D\rightarrow \infty$
   for VMPS and NRG as function of the dimension $D$ of states kept.
}
\label{fig_Eflow}%
\end{figure}

\emph{Application to Kondo model:---} Let us now illustrate above
strategies by applying them to the Kondo model. Since the Hamiltonian
in \Eq{eq.Hamilton} couples $\uparrow$ and $\downarrow$ band
electrons only via the impurity spin, it is possible (see also
\cite{Raas,Saberi08}) to ``unfold'' the semi-infinite Wilson chain
into an infinite one, with $\uparrow$ band states to the left of the
impurity and $\downarrow$ states to the right, and hopping amplitudes
decreasing in both directions as $\Lambda^{-|n|/2}$.  Since the left
and right end regions of the chain, which describe the model's
low-energy properties, are far apart and hence interact only weakly
with each other (analyzed quantitatively in terms of mutual
information in Fig.~\ref{fig_Eflow}b), the effective Hamiltonian for
these low energies will be of the form ${\cal H}^{\rm eff}_\uparrow
\otimes 1 \hspace{-1.2mm} 1_\downarrow + 1 \hspace{-1.2mm} 1_\uparrow
\otimes {\cal H}^{\rm eff} _\downarrow$. Due to the
symmetry of the Kondo coupling, ${\cal H}^{\rm eff}_\uparrow $ and
${\cal H}^{ \rm eff}_\downarrow$ have the same eigenspectrum for $n
\gg 1$, such that the fixed point spectrum is already well-reflected
by analyzing either one, as illustrated in Fig.~\ref{fig_Eflow}(a).
   Note that for a direct comparison with NRG, the spin chains can be
\emph{recombined} within VMPS \cite{Saberi08}. The resulting
standard energy flow diagram presented in panel (a) for VMPS
and NRG, respectively, show excellent agreement for low energies for
all $n$ including the fixed point spectrum.

The dimensions of the effective Hilbert spaces needed for VMPS and NRG
to capture the low energy properties (here energy resolution better
than $T_K$) are roughly related by $D_{\rm MPS} \sim \sqrt{ D_{\rm
NRG}}$ \cite{Saberi08}), implying significant computational gain
with VMPS, as calculation time scales as $D^3$ for both. Indeed,
Fig.~\ref{fig_Eflow}(e) shows that VMPS has three orders of
magnitude of better precision for the \emph{same} $D$.  More
generally, if the impurity couples to $n$ electronic bands
(channels), the Wilson chain may be unfolded into a star-like
structure of $2n$ branches, with $D_{\rm MPS} \sim D_{\rm
NRG}^{1/2n}$.  This implies that for maintaining a desired precision
in going from 1 to $n$ channels, $D_{\rm MPS}$ will stay roughly
constant, and calculation time for all sites other than the impurity
will scale merely \emph{linearly} with the number of channels.
Whether the chains can be unfolded in practice can easily be
established by checking whether or not the correlation between
them, characterized e.g. in terms of mutual information, decays
rapidly with increasing $n$ (cf. Fig.~\ref{fig_Eflow}b and caption).

\begin{figure}[t]
\includegraphics[width=.95\linewidth]{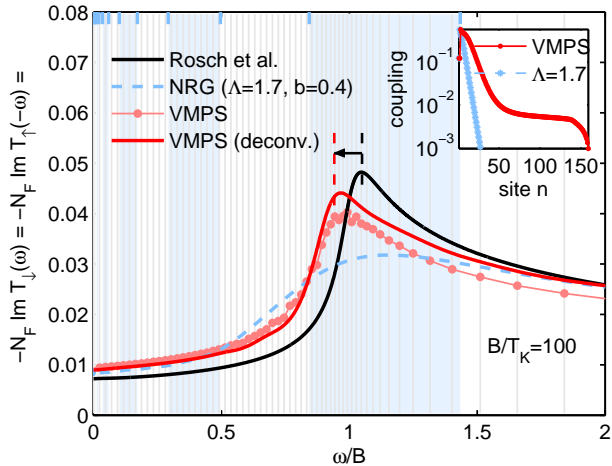}
\caption{
   (Color online) Impurity spectral function for the Kondo model 
   $-N_F\,\Im\mathrm{m}\,T_\mu(\omega)
    = J^2 \langle\langle
      \mathcal{O}_\mu^\dagger \vert \mathcal{O}_\mu
      \rangle\rangle_\omega$
   for $B \gg T_K$, where $\mathcal{O}_\mu \equiv
   \mathbf{S}\cdot\mathbf{\sigma}_{\mu\mu'}c_{\mu'}^\dagger$
   and $N_F$ is the density of states at the Fermi energy, calculated
   with VMPS (dots: raw data, red solid: deconvoluted), 
   NRG (dashed) and perturbative (black solid) \cite{Rosch03}.
      According to \cite{Rosch03}, the peak of the perturbative
   result should be shifted in $\omega$ by $-B/2\log(B/T_K)$
   (arrow).  NRG and VMPS discretization intervals are indicatd by
   blue shaded areas and gray vertical lines, respectively.
      The inset shows the hopping amplitudes corresponding to
   standard (blue dashed) and adapted (red solid) discretization schemes.
      The required Lorentzian broadening $\eta$ of the
   VMPS data smears out sharper features. Deconvolution (targeting
   with adaptive spline) together with subsequent Gaussian
   broadening was applied to obtain the red solid line \cite{EPAPS}.
}
\label{Fig_Spectral}
\end{figure}

\emph{Adaptive discretization:---}
   Through its variational
character, VMPS does not rely on logarithmic discretization crucial
for NRG. The potential of greatly enhanced energy resolution using
VMPS is already indicated by the $\Lambda=1.2$ data in
Fig.~\ref{fig_Eflow}(d). It is illustrated to full extent in
Fig.~\ref{Fig_Spectral}, showing
the splitting of the Kondo peak in the presence of a strong
magnetic field calculated using VMPS (bare: dots, deconvoluted:
red solid), standard NRG (blue dashed) and perturbatively
\cite{Rosch03} (black).
   By using a linear (logarithmic) discretization scheme for
$|\omega|\!<\!B$ ($|\omega|\!>\!B$), respectively, VMPS yields well-resolved
sharp spectral features at finite frequencies. This resolution is
out of reach for NRG, whose discretization intervals (blue shaded
intervals), even for comparatively small choice of $\Lambda=1.7$,
are much broader than the spectral features of interest.
   The line shape of our deconvoluted data (red solid line)
agrees well with the analytic RG calculation \cite{Rosch03} (black
solid line), perturbative in $1/\log(B/T_K)$.  The peak positions
agree well also after a shift in $\omega$ by $-B/2\log(B/T_K)$ of the
perturbative result suggested by \cite{Rosch03} is taken into
account.

\emph{Outlook:---} Let us finish by pointing out that
the MPS approach can readily be extended to the case of finite
temperatures by using matrix product density operators \cite{VGC}
instead of MPS, and to time-dependent problems (such as
$\mathcal{H} = \mathcal{H}(t)$ or non-equilibrium initial conditions),
by using the recently developed adaptive time-dependent DMRG
\cite{t-methods} and MPS analogues thereof \cite{VGC}.
Preliminary work in this direction was very encouraging
and will be published in the near future.

   In conclusion, the MPS approach provides a natural language for
simulating quantum impurity models. The underlying reason is
that these models, when formulated on the Wilson chain, feature
only nearest-neighbor interactions.  Their low-energy states
are thus determined mainly by their nearest-neighbor reduced
density matrices, for which very good approximations can be
obtained by suitably optimizing the set of matrices constituing
a MPS \cite{VC05}. We also
showed how these could be used for a direct variational
evaluation of Green's functions.

We gratefully acknowledge fruitful discussions with
M.~Sindel, W.~Hofstetter, G.~Uhrig and F.~Anders. This
work was supported by DFG (SFB 631, SFB-TR 12,
De 730/3-1, De 730/3-2), GIF (I-857), European projects
(Spintronics RTN, SCALA), Kompetenz\-netzwerk der
Bayerischen Staatsregierung Quanteninformation, and the
Gordon and Betty Moore Foundation (Information Science
and Technology Initiative, Caltech).
Financial support of the Excellence Cluster Nanosystems
Initiative Munich (NIM) is gratefully acknowledged.

\clearpage
                      

\section{Appendix -- Deconvolution of spectral data}

DMRG obtains spectral data from a discretized model Hamiltonian.
In order for the spectral data to be smooth, an intrinsic
frequency dependent Lorentzian broadening $\eta$ is applied during 
the calculation of the correction vector $|\chi\rangle_k$
at frequency $\omega_k$ (cf. Eq.~\ref{eq:correctionvector}),
\begin{equation}
   \delta _{\eta_k }\left( \omega - \omega_k \right) \equiv
   \frac{\eta_k}{\pi } \frac{1}{(\omega-\omega_k)^{2} + \eta_k^{2}}
   \text{.}
\label{spline.lor}
\end{equation}%
Since the original model has a continuous spectrum, the
broadening $\eta_k$ should be chosen of the order or larger than the
artificial coarse grained discretization intervals $\delta_\omega$.
Larger $\eta$ of course improves numerical convergence.  However,
since Lorentzian broadening produces longer tails than for example
Gaussian broadening, this makes it more susceptible to
pronounced spectral features closeby.
   Our general strategy for more efficient numerical treatment was
then as follows. (i) Choose somewhat larger $\eta$ ($\eta\simeq
2\delta_\omega$) throughout the calculation. (ii) Deconvolve
the raw data to such an extent that the underlying discrete
structure already becomes visible again, (iii) followed by a
Gaussian smoothening procedure which then acts more locally.
Let us describe step (ii) in more detail.

Broadening, by construction, \emph{looses} information.
Hence trying to obtain the original data from the
broadened data via deconvolution is intrinsically
ill-conditioned.  In literature there are several ways
of dealing with this problem, most prominently maximum
entropy algorithms (see \cite{Raas} and reference below).
Our approach is targeting the actual spectrual function
using the knowledge about the Lorentzian broadening used
during the VMPS calculation, combined with adaptive
spline.  Given the data $\tilde{A}\left( \omega \right)$
obtained through VMPS, let us propose the existence of
some smooth but a priori unknown target curve $A\left( \omega
\right)$, which when broadened the \emph{same} way as the
VMPS data using exactly the same $\eta_k$ via a
Lorenzian broadening kernel
\begin{equation}
   \tilde{A}_{k}\equiv \tilde{A}\left( \omega _{k}\right) =\int\limits_{-\infty
   }^{\infty }d\omega ^{\prime }~A\left( \omega ^{\prime }\right) \delta _{\eta
   _{k}}\left( \omega ^{\prime } - \omega _{k}\right) \text{,}
\label{A.kernel}
\end{equation}
reproduces the original data $\tilde{A}\left( \omega \right)$.
Direct inversion of above equation as it is is
ill-conditioned, as already mentioned, and not useful in
practice.

Let us assume the unknown target curve $A\left( \omega \right) $
is smooth and parametrized by piecewise polynomials.  Given the data
points $\omega_{k}$ with $k=1\ldots N$, the intervals in between these
values will be approximated in the spirit of adaptive spline functions
by $3^{\mathrm{rd}}$ order polynomials ($k=1\ldots N-1$)%
\begin{eqnarray}
   f_{k}\left( \omega \right) \equiv \left\{ 
   \begin{tabular}{ll}
      $a_{k} +
       b_{k}\left( \omega -\omega _{k}\right) + $ &
      $c_{k}\left( \omega -\omega _{k}\right) ^{2} + $ \\
      \hspace{.25in}$d_{k}\left( \omega -\omega _{k}\right) ^{3}
      $ & for $\omega \in \left[ \omega _{k},\omega _{k+1}\right] $ \\ 
      $0$ & otherwise.%
   \end{tabular}%
   \right. 
\label{poly:k}
\end{eqnarray}
Since spectral functions decay as $1/\omega ^{2}$ for
large $\omega$, for our purposes the ends are extrapolated
assymptotically to infinity, allowing both $1/\omega $ and
$1/\omega ^{2}$ polynomials
\begin{eqnarray}
   f_{0}\left( \omega \right)  &\equiv &\left\{ 
   \begin{tabular}{ll}
      $\frac{a_{0}}{\omega }+\frac{b_{0}}{\omega ^{2}}$
      \hspace{.25in} & $\omega \leq \omega _{1}$ \\ 
      $0$ & otherwise
   \end{tabular}
   \right.  \notag \\
   f_{N}\left( \omega \right)  &\equiv &\left\{ 
   \begin{tabular}{ll}
      $\frac{a_{N}}{\omega }+\frac{b_{N}}{\omega ^{2}}$
      \hspace{.24in} & $\omega \geq \omega _{N}$ \\ 
      $0$ & otherwise.%
   \end{tabular}%
   \right. 
\label{poly:ends}
\end{eqnarray}%
In total, this
results in $4\left( N-1\right) +2\times 2=4N$
parameters, with the target function paramatrized
piecewise as
$
   A\left( \omega \right) \equiv f \left( \omega \right) \equiv
   \sum_{k=0}^{N}f_{k}\left( \omega \right) \text{.}
$
In cases where one has not approached the assymptotic
limit yet, the ends may simply be modelled also by
\Eq{poly:k}, taking $c_0=d_0=c_N=d_N=0$. Moreover,
if information about the gradient $f'(\omega)$ is
known, it can be built in straightforwardly in the
present scheme by replacing $b_k$.

The parameters for the piecewise parametrization are
solved for by requiring the following set of conditions
\begin{description}

\item (i) \ The function $f$ should be continuous and smooth
by requiring that $f$, $f^\prime$ and $f^{\prime\prime}$
are continuous ($3N$ equations).

\item (ii) The function $f$, when broadened as in
Eq.~(\ref{A.kernel}), should reproduce the VMPS data
$\tilde{A}_{k}$ 
\begin{eqnarray}
   &&\tilde{A}_{k}^c \equiv
   \sum_{k^{\prime}=0}^{N}\int\limits_{\omega
   _{k^{\prime }}}^{\omega _{k^{\prime }+1}}d\omega ^{\prime }~f_{k^{\prime
   }}\left( \omega ^{\prime }\right) \frac{\eta _{k}/\pi }{\left( \omega
   ^{\prime }-\omega _{k}\right) ^{2}+\eta _{k}^{2}}  \label{Ak.ints} \\
   &&\tilde{A}_{k}-\tilde{A}_k^c=p_{k}r_{k} \label{rk.def}
\end{eqnarray}%
where $r_{k}\equiv f_{k}^{\left( 3\right)}(\omega_{k})
-f_{k-1}^{\left( 3\right) }(\omega_{k})$ and
$\omega_0\equiv -\infty$, $\omega_{N+1}\equiv+\infty$
($N$ equations).

\end{description}
In the spirit of adaptive spline, the
third derivative of the piecewise polynomials is no
longer required to be continuous. Its jump $r_k$ is
set proportional to the change in $\tilde{A}_{k} -
\tilde{A}_k^c$ introducing the additional
prespecified parameter set $p_k$, kept small for our
purposes (note that enforcing the strict equality
$\tilde{A}_k^c = \tilde{A}_k$ by setting $p_k=0$
would result in an ill-conditioned problem).

If intervall spacings specified by $\omega_{k}$ are
nonuniform, the $p_{k}$ have to be adapted accordingly.  For
this paper we used $p_k=p\cdot(\omega_{k+1}-\omega_k)^\alpha$ with
$p$ on the order of $10^{-6}$ and $\alpha \simeq 1$.
With $p_k$ fixed, Eqs.~(\ref{Ak.ints}) and
(\ref{rk.def}) determine all spline parameters uniquely
in terms of the original VMPS data $\tilde A_k$.
The integrals emerging out of Eq.~(\ref{Ak.ints})
can all be evaluated analytically.
The final inversion of
\Eq{Ak.ints} to obtain the parameters for $f(\omega)$
is well-behaved for small but finite $p$, small
enough to clearly sharpen spectral features.\\[2ex]
\noindent {\bf Further reading}\\[1ex]
W.H. Press, S.A. Teukolsky, W.T. Vetterling, B.P. Flannery,
Numerical Recipies in C, 2nd ed., Cambridge University
Press, Cambridge (1993).


\begin{thebibliography}{99}
\bibitem{othermethods} Compared to other techniques for calculating
Green's functions
\cite{Costi,Hofstetter,Raas,Nishimoto,GreensDMRG1,GreensDMRG2,GreensDMRG3},
the VMPS approach proposed here has the advantage that it is
variational, and hence in principle optimal within the set of MPS.
It is more efficient than the continued fraction
method \cite{GreensDMRG1}, the correction vector method
\cite{GreensDMRG2} and dynamical DMRG \cite{GreensDMRG3}, because
each of these methods require several states to be calculated
simultaneously,
thus requiring larger $D$ for the same precision.
\end{thebibliography}
\end{document}